\date{}
\documentclass[aps,pra,showpacs,twocolumn]{revtex4-1}
\usepackage{graphicx,psfrag,amsmath,amssymb,amsfonts,latexsym,color,dcolumn}

\begin{document}

\title{The proximity force approximation for the Casimir energy as a derivative expansion}

\author{C\'esar D. Fosco$^{1,2}$}
\author{Fernando C. Lombardo$^3$}
\author{Francisco D. Mazzitelli$^{1,3}$}

\affiliation{$^1$ Centro At\'omico Bariloche,
Comisi\'on Nacional de Energ\'\i a At\'omica,
R8402AGP Bariloche, Argentina}
\affiliation{$^2$ Instituto Balseiro,
Universidad Nacional de Cuyo,
R8402AGP Bariloche, Argentina}
\affiliation{$^3$ Departamento de F\'\i sica {\it Juan Jos\'e
 Giambiagi}, FCEyN UBA, Facultad de Ciencias Exactas y Naturales,
 Ciudad Universitaria, Pabell\' on I, 1428 Buenos Aires, Argentina - IFIBA}

\date{today}

\begin{abstract} 
The proximity force approximation (PFA) has been widely used as a tool
to evaluate the Casimir force between smooth objects at small
distances. In spite of being intuitively easy to grasp, it is
generally believed to be an uncontrolled approximation. Indeed, its
validity has only been tested in particular examples, by confronting
its predictions with the next to leading order (NTLO) correction
extracted from numerical or analytical solutions obtained without
using the PFA. In this article we show that the PFA and its NTLO
correction may be derived within a single framework, as the first two
terms in a derivative expansion. To that effect, we consider the
Casimir energy for a vacuum scalar field with Dirichlet conditions on
a smooth curved surface described by a function $\psi$ in front of a
plane. By regarding the Casimir energy as a functional of $\psi$, we
show that the PFA is the leading term in a derivative expansion of
this functional. We also obtain the general form of corresponding NTLO
correction, which involves two derivatives of $\psi$. We show, by
evaluating this correction term for particular geometries, that it
properly reproduces the known corrections to PFA obtained from exact
evaluations of the energy.

\end{abstract}


\maketitle

\section{Introduction}\label{sec:intro}

In the last years, there have been important theoretical and experimental
advances in the analysis of the Casimir effect~\cite{reviews}.

 Until the recent development of theoretical methods that allowed for the exact
 evaluation of the Casimir energy for several geometries, the interaction between
different bodies has been mostly computed using the so called proximity
force approximation (PFA)~\cite{derjaguin}. This approximation, expected to
be reliable as long as  the interacting surfaces are smooth, almost
parallel, and very close, makes use of Casimir's expression for the energy
per unit area for two parallel plates at a distance $a$ apart. For the case
of a single massless scalar field and  Dirichlet conditions (the case we
deal with in this paper) it is given by:
\begin{equation}
E_{\rm pp}(a)=-\frac{\pi^2 }{1440 \, a^3}\;.
\end{equation}
The PFA then approximates the interaction between two Dirichlet surfaces
separated by a gap of spatially varying width $z$, as follows: 
\begin{equation}\label{sigma}
E_{\rm PFA}=\int_\Sigma d\sigma\, E_{\rm pp}(z) \;,
\end{equation}
where $\Sigma$ is one of the two surfaces.
Quite obviously, this formula does not take into account the
non-parallelism of the surfaces. Moreover, the result may depend on the
particular surface $\Sigma$ chosen to perform the integral. 

As the PFA was believed to be
 an uncontrolled approximation, its  accuracy has been assessed only 
 in some of the particular geometries where it was possible to 
 compute the Casimir energy numerically or analytically.
 On general grounds, denoting by $\mathcal{L}$ a typical
 length associated to the curvature of one of the surfaces (assumed much
 smaller than the curvature of the second one) and by $a$ the minimum
 distance between surfaces, one expects that:
 \begin{equation}
	 E_{\rm C}\,=\,E_{\rm
	 PFA}\left\{1+\gamma\frac{a}{\mathcal{L}}+{\cal
	 O}\left[\left(\frac{a}{\mathcal{L}}\right)^2\right]\right\},  
 \end{equation}
where $\gamma$ is a constant, whose numerical value fixes the accuracy of
the PFA in each particular geometry (the situation could be more complex,
since the corrections to PFA may contain non-analytic corrections as
\mbox{$\left(\frac{a}{\mathcal{L}}\right)^n\log\left(\frac{a}{\mathcal{L}}\right)$}).
One can write similar expressions for geometries that involve two surfaces
of similar curvature. 

In this paper we explore the following simple idea. The Casimir energy can
be thought as a functional of the shape of the surfaces of the interacting
bodies. As the PFA should be adequate for almost plane surfaces,  a
derivative expansion \cite{derivativeexp} 
of this functional should reproduce, to lowest order,
the PFA. Moreover, the terms involving derivatives of the functions that
describe the shape of the surfaces should contain the corrections to the
PFA. We will show that this is indeed the case, and that it is possible to
find a general formula to compute the first corrections to PFA for rather
arbitrary surfaces.  

Just to avoid some technical complications, we consider a massless
scalar field in the presence of a  curved surface in front of a plane. We
will assume that the quantum field satisfies Dirichlet boundary conditions
on both surfaces. Generalizations to other boundary conditions and to the
electromagnetic field will be analyzed in a forthcoming work.

This paper is organized as follows. In Section~\ref{sec:model}, we describe
the model and derive a formal expression for the Casimir energy in the
geometry described above. Then, in Section~\ref{sec:derivative}, we perform
a derivative expansion in the expression for the Casimir energy to
obtain the main result of the paper: a general formula for the interaction
energy between an arbitrary curved surface and a plane, containing up to
two derivatives of the function $\psi$ that describes the curved surface.
The leading term of the expansion corresponds to the PFA, while the term
with derivatives  is the first non trivial correction.

In Section~\ref{sec:examples}  we present some examples: a sphere, a
cylinder, or  a corrugated surface in front of a plane.  We show, by
comparing with existing analytical results, that the derivative expansion
of the Casimir energy describes correctly both the PFA and its first
correction for all of these geometries.  We also compute the derivative
expansion of the Casimir energy for geometries involving parabolic mirrors. 
Section~\ref{sec:conclusions} contains the conclusions of our work.

\section{Formal expression for the vacuum energy}\label{sec:model}

We shall consider a model consisting of a massless real scalar field
$\varphi$ in $3+1$ dimensions, coupled to two mirrors which impose
Dirichlet boundary conditions.  In our Euclidean conventions, we use
$x_0,x_1,x_2,x_3$ to denote the spacetime coordinates, $x_0$ being the
imaginary time.

The mirrors occupy two surfaces, denoted by $L$ and $R$, defined by the
equations $ x_3 = 0 $ and  $x_3 =  \psi(x_1,x_2)$,  respectively. 

Following the functional approach to the Casimir effect, we introduce 
${\mathcal Z}$, which may be interpreted as the zero temperature limit of
a partition function, for the scalar field in the presence of the two
mirrors. It may be written as follows:
\begin{equation}
	{\mathcal Z}\;=\; \int {\mathcal D}\varphi \;\delta_L(\varphi) \,
	\delta_R(\varphi) \; e^{- S_0(\varphi) } \;,
\end{equation}
where $S$ is the free real scalar field Euclidean action
\begin{equation}
S_0(\varphi) \;=\; \frac{1}{2} \, \int d^4 x \, (\partial \varphi)^2
\;,
\end{equation}
while $\delta_L$ ($\delta_R$) imposes Dirichlet boundary conditions on the
$L$ ($R$) surface.

Exponentiating the two delta functions by introducing two auxiliary fields,
$\lambda_L$ and $\lambda_R$, we obtain for ${\mathcal Z}$ an
equivalent expression:
\begin{equation}
{\mathcal Z}\;=\; \int {\mathcal D}\varphi {\mathcal D}\lambda_L
\, {\mathcal D}\lambda_R \; e^{-S(\varphi;\lambda_L,\lambda_R)} \;,
\end{equation}
with
\begin{eqnarray}
&&S(\varphi;\lambda_L,\lambda_R) =S_0(\varphi) \\&-&i 
\int d^4x \varphi(x) 
\left[ 
\lambda_L(x_\parallel) \delta(x_3) 
+ \lambda_R(x_\parallel) \delta(x_3-\psi({\mathbf x_\parallel})) \right]
\nonumber \end{eqnarray}
where we have introduced the notations $x_\parallel\equiv (x_0,x_1,x_2)$
and ${\mathbf x_\parallel} \equiv (x_1,x_2)$.

Integrating out $\varphi$, we see that ${\mathcal Z}_0$,
corresponding to the field $\varphi$ in the absence of boundary conditions
factors out, while the rest becomes an integral over the auxiliary fields:
\begin{equation}
{\mathcal Z}={\mathcal Z}_0  
\int {\mathcal D}\lambda_L  {\mathcal D}\lambda_R  
e^{-\frac{1}{2} \int d^3x_\parallel \int d^3y_\parallel 
\sum_{\alpha,\beta}\lambda_\alpha(x_\parallel)  
{\mathbb T}_{\alpha\beta} 
\lambda_\beta(y_\parallel)},
\end{equation}
where $\alpha,\,\beta=L,R$ and we have introduced the objects:
\begin{align}
{\mathbb T}_{LL}(x_\parallel,y_\parallel) &=\langle
x_\parallel,0|(-\partial^2)^{-1} |y_\parallel,0\rangle \\
{\mathbb T}_{LR}(x_\parallel,y_\parallel) &=\langle
x_\parallel,0|(-\partial^2)^{-1}
|y_\parallel,\psi({\mathbf y_\parallel})\rangle \\
{\mathbb T}_{RL}(x_\parallel,y_\parallel) &=\langle
x_\parallel,\psi({\mathbf x_\parallel})|(-\partial^2)^{-1} 
|y_\parallel,0\rangle \\
{\mathbb T}_{RR}(x_\parallel,y_\parallel) &=\langle
x_\parallel,\psi({\mathbf x_\parallel})|(-\partial^2)^{-1} 
|y_\parallel,\psi({\mathbf y_\parallel})\rangle 
\end{align}
where we use a ``bra-ket'' notation to denote matrix elements of operators,
and $\partial^2$ is the four-dimensional Laplacian. Thus, for example,
\begin{equation}
	\langle x|(-\partial^2)^{-1} |y\rangle \,=\, \int
	\frac{d^4k}{(2\pi)^4} \, \frac{e^{i k \cdot (x-y)}}{k^2}\;.
\end{equation}
The vacuum energy of the system, $E_{\rm vac}$, subtracting the zero-point
energy of the free field (contained in ${\mathcal Z}_0$), is: 
\begin{equation}\label{eq:evac}
	E_{\rm vac}\;=\; \lim_{T \to \infty} \big(\frac{\Gamma}{T}\big)\;=\; 
	\frac{1}{2 T} {\rm Tr}  \log {\mathbb T} \;,
\end{equation}
where $T$ is the extent of the time dimension (or $\beta^{-1}$, in the
thermal partition function setting), $\Gamma \equiv - \log\frac{\mathcal
Z}{{\mathcal Z}_0}$ and the trace is meant to act on both
discrete and continuous indices.

Note that $E_{\rm vac}$ still contains `self-energy' contributions, due 
to the vacuum distortion produced by each mirror, even when the other is
infinitely far apart. This piece (irrelevant to the force between mirrors)
shall be subtracted, in order to obtain a finite Casimir energy, 
in the calculations below. 

\section{Derivative expansion}\label{sec:derivative}

We present here a derivation of the first two terms in a derivative
expansion of the Casimir energy for the system defined in the previous
section.

To that end, and for calculational purposes, it is convenient 
to consider first a simplified situation: we split $\psi$ into 
two components, 
\begin{equation}
	\psi({\mathbf x_\parallel}) \;=\; a \,+\, \eta({\mathbf
	x_\parallel}) \;,
\end{equation}
where $a$ (assumed to be greater than zero), is the spatial average of
$\psi$, and therefore a constant, whereas $\eta$ contains the varying piece
of $\psi$. The simplified case amounts to expanding up to the second order in
$\eta$. Since the derivatives of $\psi$ equal the derivatives of $\eta$,
to find the terms with up to two derivatives of $\psi$, it is sufficient to
expand $\Gamma$ up to the second order in $\eta$, keeping up to the second
order term in an expansion in derivatives:
\begin{equation}\label{eq:expansion1}
	\Gamma(a,\eta)\;=\; \Gamma^{(0)}(a) \;+\;\Gamma^{(1)}(a,\eta) \;+\;
 \Gamma^{(2)}(a,\eta) \;+\;\ldots
\end{equation}
where the index denotes the order in derivatives. Each term will be a
certain coefficient times the spatial integral over ${\mathbf x_\parallel}$
of a local term, depending on $a$ and derivatives of $\eta$.

So far this is a perturbative expansion in $\eta$ and its derivatives. 
However, to the same order in derivatives, it is quite 
straightforward to include the corrections which are of the same order
in derivatives but of arbitrary order in $\eta$. 
Indeed, to do this, in the terms obtained in (\ref{eq:expansion1}),  
one just has to replace  $a$ by $\psi$ and also $\eta$ by $\psi$, before
performing the spatial integrals.
This procedure accounts for all the terms of higher order in $\eta$, and the
same order in derivatives, that contribute to the respective order the
derivative expansion. Formally, this procedure may be represented as
follows:
\begin{equation}\label{eq:expansion2}
	\Gamma^{(l)}(\psi) \;=\; \Gamma^{(l)}(a,\eta)\big|_{a \to \psi, \eta \to \psi}  
\end{equation}
for each term in (\ref{eq:expansion1}).

Let us calculate the different terms in the derivative expansion for
$\Gamma$, following this procedure.

Expanding first the matrix ${\mathbb T}$ in powers of $\eta$
\begin{equation}
{\mathbb T}={\mathbb T}^{(0)}+{\mathbb T}^{(1)}+{\mathbb T}^{(2)}+\ldots\, ,
\end{equation}
we obtain $\Gamma \,=\,
\Gamma^{(0)} + \Gamma^{(1)}+ \Gamma^{(2)}\,+\,\ldots$, where
\begin{eqnarray}
\Gamma^{(0)}&=&\frac{1}{2} {\rm Tr}  \log {\mathbb T}^{(0)}\nonumber\\
\Gamma^{(1)}&=&\frac{1}{2} {\rm Tr}  \log \left[({\mathbb T}^{(0)})^{-1}
{\mathbb T^{(1)}}\right]\nonumber\\
\Gamma^{(2)}&=&\frac{1}{2} {\rm Tr}  \log \left[({\mathbb T}^{(0)})^{-1}{\mathbb T}^{(2)}\right]
\nonumber\\
&&-\frac{1}{4} {\rm Tr}  \log \left[ ({\mathbb T}^{(0)})^{-1}{\mathbb T}^{(1)}  
({\mathbb T}^{(0)})^{-1}{\mathbb T}^{(1)}
\right]\;,
\end{eqnarray}
where, in $\Gamma^{(l)}$, we need to keep up to $l$ derivatives of
$\eta$.

The zeroth-order term is thus simply obtained by replacing first $\psi$ by a
constant, $a$, and then subtracting the contribution corresponding to $a
\to \infty$, to get rid of the divergent self-energies. This yields,
\begin{equation}
	\Gamma^{(0)}(a) \;=\; \frac{1}{2} \, {\rm Tr} \log \big[ 1 - 
	(T_{LL}^{(0)})^{-1} T_{LR}^{(0)}
	(T_{RR}^{(0)})^{-1} T_{RL}^{(0)} \big]
\end{equation}
where the $T_{\alpha\beta}^{(0)}$ elements are identical to the ones on
would have for the two flat parallel mirrors at a distance $a$ apart. As
mentioned above, we have then to replace $a$ by $\psi$ at the end of the
calculation.  
After evaluating the trace, we obtain:
\begin{equation}
\Gamma^{(0)}\,=\, \frac{T}{2} \, \int d^2{\mathbf x_\parallel}\int\frac{d^3k_\parallel}{(2\pi)^3}
\log[1-e^{-2k_\parallel a}]\, .
\end{equation}
We then replace $a \to \psi$ to extract the zeroth order Casimir energy,
\begin{eqnarray}
E_{\rm vac}^{(0)} &=& \frac{1}{2} \, \int d^2{\mathbf x_\parallel}\int\frac{d^3k_\parallel}{(2\pi)^3}
\log[1-e^{-2k_\parallel\psi({\mathbf x_\parallel})}] \nonumber \\
&=& -\frac{\pi^2}{1440} \int d^2{\mathbf x_\parallel}\; \frac{1}{\psi({\mathbf x_\parallel})^3}  ,
\end{eqnarray}
which equals the PFA approximation to the vacuum energy.

The first order term in the derivative expansion $\Gamma$ 
vanishes identically, while for the second order one we have two
contributions:
\begin{equation}
	\Gamma^{(2)} \;=\; \Gamma^{(2,1)} \,+\, \Gamma^{(2,2)} 
\end{equation}
where,
\begin{equation}
\Gamma^{(2,1)} \,=\, \frac{1}{2} {\rm Tr}  \log \left[({\mathbb
T}^{(0)})^{-1}{\mathbb T}^{(2)}\right] 
\end{equation}
and
\begin{equation}
\Gamma^{(2,2)} \,=\,-\frac{1}{4} {\rm Tr}  \log \left[ ({\mathbb
T}^{(0)})^{-1}{\mathbb T}^{(1)}  ({\mathbb T}^{(0)})^{-1}{\mathbb T}^{(1)}
\right] \;,
\end{equation}
where we have to keep up to two derivatives of $\eta$.

The form of those terms can be obtained in a quite straightforward fashion;
indeed, we first note that, in Fourier space, and before expanding to
second order in momentum (derivatives), they have the structure:
\begin{equation}
	\Gamma^{(2,j)}\;=\; \frac{T}{2} \, \int \frac{d^2k}{(2\pi)^2}
	f^{(2,j)}({\mathbf k}) \, |\tilde{\eta}({\mathbf k})|^2 
\end{equation}
($j=1,2$), where ${\mathbf k}= (k_1,k_2)$, $\tilde{\eta}$ is the Fourier
transform of $\eta$, and the $f^{(2,j)}$ kernels are the $k_0\to 0$ (i.e.,
static) limits 
of:
\begin{eqnarray}
	f^{(2,1)}(k) &=& - \int \frac{d^3p}{(2\pi)^3} \frac{|p| \, |p+k|}{1
	- e^{- 2 |p + k| a}} \nonumber\\
	f^{(2,2)}(k) &=& - \int \frac{d^3p}{(2\pi)^3} 
	\frac{|p| |p+k| e^{-2 |p+k| a} ( 1 + e^{-2 |p| a}) }{(1 - e^{- 2 |p| a}) (1 - e^{- 2 |p + k| a})}\nonumber .
\end{eqnarray}
Besides, we need to subtract an $a$-independent self-energy contribution,
obtained by taking $a\to \infty$ in the expressions above. Putting together
the two terms above, and subtracting the $a \to \infty$ limit, 
the total contribution to $\Gamma^{(2)}$ adopts the form:
\begin{equation}
	\Gamma^{(2)}\;=\; \frac{T}{2} \,\int \frac{d^2k}{(2\pi)^2}
	f^{(2)}({\mathbf k}) \, |\tilde{\eta}({\mathbf k})|^2 
\end{equation}
with:
\begin{equation}
f^{(2)}(k) \;=\; - 2 \int \frac{d^3p}{(2\pi)^3} 
\frac{|p| \,|p+k|}{(1 - e^{- 2 |p| a}) (e^{2 |p+k| a} - 1)}
\end{equation}
where we just need to extract its ${\mathbf k}^2$ term in a 
Taylor expansion at zero momentum. Namely $f^{(2)}({\mathbf k}) \;\simeq\; \chi  \, {\mathbf k}^2$,
where
\begin{eqnarray}
	\chi &=&  \frac{1}{2} \,  \big[\frac{\partial^2 f^{(2)}(k)}{\partial k^2}\big]_{k
	\to 0} \nonumber\\
	&=& - \, \int \frac{d^3p}{(2\pi)^3} \, \frac{|p|}{(1 - e^{- 2|p|a})} \, 
	\lim_{k \to 0} \, \frac{\partial^2 }{\partial k^2}
\Big[\frac{|p+k|}{(e^{2 |p+k| a} - 1)} \Big] \nonumber .
\end{eqnarray}
The resulting integral may be exactly calculated,
\begin{equation}
	\chi \,=\, - \frac{\pi^2}{1080 \, a^3} \;.
\end{equation}
Thus,
\begin{eqnarray}
\Gamma^{(2)}(a,\eta) &=&- \frac{T}{2} 
\frac{\pi^2}{1080} 
\int \frac{d^2k}{(2\pi)^2}
\frac{{\mathbf k}^2}{a^3} \, |\tilde{\eta}({\mathbf k})|^2 \nonumber\\
&=&-\frac{T}{2} \frac{\pi^2}{1080} \int d^2{\mathbf x_\parallel}\,
\frac{1}{a^3} \, (\partial_\alpha \eta)^2 \;,
\end{eqnarray}
where, to obtain the second order contribution in derivatives to the vacuum
energy, we  need to replace $a \to \psi$,  $\eta \to \psi$, and cancel the
$T$ factor, obtaining:
\begin{equation}
E^{(2)}_{\rm vac}=\frac{\Gamma^{(2)}(\psi)}{T} \,= \, -\frac{1}{2} 
\frac{\pi^2}{1080} \int d^2{\mathbf x_\parallel}\,
\frac{(\partial_\alpha \psi)^2}{\psi^3} \;,
\end{equation}
where the index $\alpha$ runs from $1$ to $2$.

Putting together the terms up to second order, the expression for the
energy becomes:
\begin{eqnarray}
E_{\rm DE} &\equiv &
E_{\rm vac}^{(0)}+E_{\rm vac}^{(2)}\nonumber \\ &=&- \frac{\pi^2}{1440}\int d^2{\mathbf x_ \parallel} \;
\frac{1}{\psi^3}\left[1+\frac{2}{3}(\partial_\alpha\psi)^2\right].
\label{final}\end{eqnarray}
This is the main result of this paper.
The first term  is the PFA for the Casimir energy. The second term contains
the first non-trivial correction to PFA for an arbitrary surface. We could
have guessed the form of both terms in the final formula by using
dimensional and symmetry arguments. The global factor could also be
determined by considering the particular case of parallel plates.
Therefore,  the calculation presented above, besides confirming the general
arguments,  provides the relative weight between both terms, which turns
out to be $2/3$, regardless of the form of the surface.

\section{Examples}\label{sec:examples}

We provide here some applications of the general formula for the Casimir interaction energy.

\subsection{A corrugated surface in front of a plane}
Let us first consider a corrugated surface in front of a plane. For
simplicity we assume sinusoidal corrugations in the  direction of $x_1$
\begin{equation}
\psi(x_1)= a +\epsilon\sin\left(\frac{2\pi x_1}{\lambda}\right),
\end{equation}
where $a$ is the mean distance to the flat surface, $\epsilon$ is the
amplitude, and 
$\lambda$ the wavelength of the corrugation. We assume a square plane of 
side $L$, which is much larger than any other length in the problem.

The derivative expansion for the Casimir energy is given by
\begin{eqnarray}
E_{\rm DE}&=&- \frac{\pi^2}{1440}\left[\int d^2{\mathbf x_\parallel}\;\frac{1}{(a +\epsilon \sin\frac{2\pi x_1}{\lambda})^3}\right. \nonumber \\
&\times & \left. \left(1+\frac{2}{3}
\left(\frac{2\pi}{\lambda}\right)^2\epsilon^2\cos^2 \frac{2\pi
x_1}{\lambda} \right)\right] \;.
\end{eqnarray}
In this case, the derivative expansion is an expansion in  powers of
$a/\lambda$ and $\epsilon/\lambda$, i.e.  $\lambda$ is largest relevant distance in the
problem. In order to compare with previous results in the literature
\cite{golest}, we will further assume that $\epsilon\ll a$. In this limit
we obtain
\begin{equation}
E_{\rm DE}\simeq - \frac{\pi^2L^2}{1440 a^3}\left[1+3\left(\frac{\epsilon}{a}\right)^2+\frac{4\pi^2}{3}
\left(\frac{\epsilon}{\lambda}\right)^2\right]\, .
\label{EDE corr}
\end{equation}
 This expression coincides with the small $a/\lambda$ expansion of the result
 obtained in Ref.\cite{golest}. Indeed, in that work the interaction energy was written as
 \begin{equation}
 \frac{E_{\rm vac}}{L^2}= - \frac{\pi^2}{1440 a^3}-\frac{\epsilon^2}{a^5}G_{\rm TM}\left(\frac{a}{\lambda}\right)\, ,
 \label{golest}
\end{equation}
where $G_{\rm TM}(x)$ can be written in terms of  Polylogarithm functions
\cite{golest2}. One can readily compute the small argument expansion of $G_{TM}$ to obtain
\begin{equation}
G_{\rm TM}(x)\simeq  \frac{\pi^2}{480}+\frac{\pi^4x^2}{1080}\, .
\end{equation}
After inserting this expansion into Eq. (\ref{golest}),  the result coincides with the derivative
expansion Eq. (\ref{EDE corr}).

\subsection{A sphere in front of a plane}

We now consider a sphere of radius $R$ at a distance $a$ from  a plane. 
The evaluation of the Casimir energy in the electromagnetic case for this configuration has 
been performed in Refs. \cite{17, 18}, while the evaluation for scalar fields has been previously
reported in Ref. \cite{11}. See also \cite{30,Bordag2010} for asymptotic expansions in the scalar and
electromagnetic cases near
the proximity limit.

For this geometry, we expect the derivative expansion to be adequate in the
 limit $a\ll R$.  It is worth noting that the surface of the sphere cannot
be described by a single valued function $x_3=\psi(x_1,x_2)$.  Note that
even if we consider an hemisphere, the derivatives of $\psi$ will be
divergent on the equator.  For these reasons, the derivative expansion will
not converge. In spite of this, we will see that it still gives
quantitative adequate results even beyond the lowest order approximation.

In order to avoid these problems, we will consider only the region of the
sphere which is closer to the plane. This is the usual approach when
computing the Casimir energy using the  PFA. The final result will not
depend on the part of the sphere considered. Denoting by $(\rho,\varphi)$
the polar coordinates in the $(x_1,x_2)$ plane the function $\psi$ reads
\begin{equation}
\psi(\rho)=a+R\left(1-\sqrt{1-\frac{\rho^2}{R^2}}\right)\, .
\label{funsphere}
\end{equation} 
This function describes an hemisphere when $0\leq \rho\leq R$. As mentioned above,
the derivative expansion will be well defined if we restrict the integrations to the region
$0\leq\rho\leq\rho_{\rm M}< R$. 

Inserting this expression for $\psi$ into the derivative expansion for the Casimir energy, one can 
perform explicitly
the integrations and obtain an analytic expression $E_{\rm DE}(\rho_{\rm M},a,R)$. 
We do not present this rather long expression here, but only the leading terms
in an  expansion in powers of $a/R$, which is  given by 
\begin{eqnarray}
E_{\rm vac}^{(0)} &\simeq&- \frac{\pi^3}{1440}\frac{R}{a^2}\Big[ 1-\frac{a}{R}\Big]\label{E0sphere}\\
E_{\rm vac}^{(2)} &\simeq& -\frac{\pi^3}{1080\, a}\, ,
\end{eqnarray}
and therefore
\begin{equation}
E_{\rm DE}\simeq - \frac{\pi^3}{1440}\frac{R}{a^2}\left(1+\frac{1}{3}\frac{a}{R}\right).
\end{equation}
It noteworthy that, up to this order, the result does not depend on $\rho_{\rm M}$. Moreover, the result
is in agreement with the asymptotic expansion obtained from the exact formula for this configuration
\cite{30},  and with the former numerical evaluation in \cite{gies2003}.

It is interesting to remark that $E_{\rm vac}^{(0)}$ includes part of the
next to leading order corrections. It is correct to keep the second term in
Eq. (\ref{E0sphere}) only when the contribution coming from   $E_{\rm
vac}^{(2)}$ is also taken into account.

\subsection{A cylinder in front of a plane}

Let us now consider a cylinder of radius $R$ and length $L\gg R$ at a distance $a$ from  a plane. The Casimir energy 
for this configuration was first evaluated in the PFA in Ref.\cite{27}.  The exact result was first derived in Ref.\cite{Emig2006}.
The caveats mentioned in the above subsection also apply for this geometry. We will consider the function $\psi$ given by
\begin{equation}
\psi(x_1)=a+R\left(1-\sqrt{1-\frac{x_1^2}{R^2}}\right)\, ,
\label{funcyl}
\end{equation} 
with $-x_{\rm M}<x_1<x_{\rm M}<R$ in order to cover the part of the cylinder which is closer to the plane. The calculation is similar to the previous case, and the final result is
\begin{equation}
E_{\rm DE}\simeq - \frac{\pi^3L}{1920\sqrt 2}\frac{R^{1/2}}{a^{5/2}}\left(1+\frac{7}{36}\frac{a}{R}\right).
\end{equation}
Once more, up to this order, the result does not depend on $x_{\rm M}$. Moreover, it is in agreement with the 
asymptotic expansion obtained from the exact formula for the cylinder-plane geometry and numerical findings 
\cite{Bordag, 29bis, PRDnum, 29}.

\subsection{A parabolic cylinder in front of a plane}

We compute here  the Casimir interaction energy between a parabolic cylinder of length $L$ in front of a plane.  
The surface is defined by the function
\begin{equation}
\psi(x_1)=a+\frac{x_1^2}{2R}\, ,
\label{funparcyl}
\end{equation} 
with $-x_{\rm M}<x_1<x_{\rm M}<R$.   Once more, we only consider the portion of the curved surface which is closer to the plane
(note that the functions defining the cylinder Eq.(\ref{funcyl}) and the parabolic 
cylinder Eq.(\ref{funparcyl}) coincide up to first order in $x_1/R$). The integrations needed to compute the derivative expansion of the Casimir energy are very simple. Expanding the result
in powers of $a/R$ we obtain
\begin{equation}
E_{\rm DE}\simeq - \frac{\pi^3L}{1920\sqrt{2}}\frac{R^{1/2}}{a^{5/2}}\left(1+\frac{4}{9}\frac{a}{R}\right).
\end{equation}
The final answer is independent of $x_{\rm M}$ and the leading order coincides with  that of the cylinder
in front of a plane.
 
\subsection{A paraboloid in front of a plane}

As a final example we consider a paraboloid, defined by
\begin{equation}
\psi(\rho)=a+\frac{\rho^2}{2R}\, ,
\label{funparabol}
\end{equation} 
with $0<\rho<\rho_{\rm M}<R$, in front of a plane. 

The approximation for the vacuum energy reads
\begin{equation}
E_{\rm DE}\simeq - \frac{\pi^3}{1440}\frac{R}{a^2}\left(1+\frac{4}{3}\frac{a}{R}\right).
\end{equation}
As in all the previous examples, the result does not depend on the region of integration defined by $\rho_{\rm M}$. Moreover,
the leading order is equal to that of the sphere in front of a plane, as expected from the fact that the functions 
describing both surfaces Eqs. (\ref{funsphere}) and (\ref{funparabol}) coincide in the region closer to the plane.

\section{Conclusions}\label{sec:conclusions}
We have shown that the PFA can be thought of as akin to a derivative expansion
of the Casimir energy with respect to the shape of the surfaces. Our main result, given in Eq. (\ref{final}),
shows that the lowest order (the ``effective potential'')  reproduces the PFA. Moreover, when the 
first non trivial correction containing two derivatives of $\psi$ is also included, the general formula
gives the NTLO correction to PFA for a general surface.

Several remarks are in order: to begin with, at least for the surfaces considered in this paper,
the PFA becomes a well defined and controlled approximation scheme: 
the leading corrections are small when $\vert \partial_\alpha\psi\vert \ll 1$ or, in other words,
when the curved surface is almost parallel to the plane. Higher order corrections will be negligible
when, in addition to this condition, the scale of variation of the shape of the surface is
much larger than the local distance between surfaces. It is also clear that the corrections
to PFA only contains local information about the geometry of the surface, and 
does not include correlations between different points of the surface.

Although we applied our general result to the case of a cylinder and a
sphere in front of a plane, these geometries present additional
complications, because they cannot be described by a single
function $\psi$. Moreover, the derivatives of $\psi$  diverge when the
surface becomes perpendicular to the plane, and therefore it is clear that the derivative
expansion will not converge.  In spite of this, it is remarkable that Eq. (\ref{final}) describes
the interaction energy for these configurations including the first non trivial correction to PFA. 
Strictly speaking, for these geometries we are computing the interaction energy between a plane  and a large curved surface which, in the region closest to the plane, has a cylindrical or spherical shape.

We expect the  main idea presented in this paper  to be generalizable in several directions, as for instance for a scalar field satisfying Neumann or Robin boundary conditions, and also to the 
electromagnetic field satisfying perfect conductor boundary conditions on the surfaces. In all 
these cases, we expect the derivative expansion 
to be of the form 
\begin{equation}
E_{\rm DE} = - \frac{\pi^2}{1440}\int d^2{\mathbf x_ \parallel} \;
\frac{1}{\psi^3}\left[\beta_1+\beta_2(\partial_\alpha\psi)^2\right]\, ,
\label{final2}\end{equation}
where the constants $\beta_i$ will depend on the kind of fields and boundary conditions considered.

Other interesting generalizations would be to consider two curved surfaces, and the case
of imperfect boundary conditions. Moreover, as the applications 
of the PFA are not restricted to the Casimir energy,  the derivative expansion could also be 
useful to compute gravitational \cite{gravitational}, electrostatic \cite{elec} or even nuclear forces \cite{Blocki}.

\section*{Acknowledgements}
This work was supported by ANPCyT, CONICET, UBA and UNCuyo.

\end{document}